\documentclass[5p,times,twocolumn]{elsarticle}
\journal{Physics Letters B}

\usepackage{amssymb,amsmath}
\usepackage{slashed}
\usepackage{hyperref}
\hypersetup{
    pdfauthor={Kenshi Kuroki},
    colorlinks=true,
    allcolors=blue
    }
\usepackage{comment}
\allowdisplaybreaks[3]

\begin{document}

\begin{frontmatter}

\title{
Tensor-polarized twist-3 parton distribution functions \texorpdfstring{$f_{LT} (x)$}{fLT(x)}\\
for the spin-1 deuteron by using twist-2 relations
}

\author[QMRC,SCNT,KEK]{S. Kumano}
\ead{kumanos@impcas.ac.cn}
\author[QMRC,SCNT]{Kenshi Kuroki\corref{cor}}
\ead{k-kuroki@impcas.ac.cn}
\cortext[cor]{Corresponding author}

\affiliation[QMRC]{
    organization={Quark Matter Research Center, Institute of Modern Physics, Chinese Academy of Sciences},
    city={Lanzhou},
    postcode={73000},
    country={China}
}
\affiliation[SCNT]{
    organization={Southern Center for Nuclear Science Theory, Institute of Modern Physics, Chinese Academy of Sciences},
    city={Huizhou},
    postcode={51600},
    country={China}
}
\affiliation[KEK]{
    organization={KEK Theory Center, Institute of Particle and Nuclear Studies, KEK},
    city={Tsukuba},
    postcode={305-0801},
    country={Japan}
}

\begin{abstract}
Tensor-polarized twist-3 parton distribution functions (PDFs) $f_{LT} (x)$ are calculated for the spin-1 deuteron by using twist-2 relations, which are similar to the Wandzura-Wilczek relation and the Burkhardt-Cottingham sum rule in the spin-1/2 nucleon, together with tensor-polarized twist-2 PDFs $f_{1LL}(x)$.
The PDFs are shown for $f_{LT} (x)$ at $Q^2 =2.5$~GeV$^2$, where the tensor-polarized PDFs $f_{1LL}(x)$ are provided. The $x$-dependence of $f_{LT} (x)$ is similar to $f_{1LL}(x)$, and the magnitude of $f_{LT} (x)$ is roughly of the order of $f_{1LL}(x)$. 
In experiments at the Thomas Jefferson National Accelerator Facility (JLab), higher-twist effects could be sizable because $Q^2$ values are not very large in comparison with the hadronic scale of 1~GeV$^2$.
Therefore, the JLab experiments could provide a good opportunity to investigate the twist-3 distributions $f_{LT} (x)$ in addition to the twist-2 ones $f_{1LL} (x)$.
Furthermore, these tensor-polarized PDFs could be investigated at future Electron-Ion Colliders (EICs) and hadron accelerator facilities such as the Fermi National Accelerator Laboratory (Fermilab), the Nuclotron-based Ion Collider fAcility (NICA), and the Large Hadron Collider (LHC).
\end{abstract}


\end{frontmatter}

\section{Introduction}
\label{introduction}

Polarized structure functions of the spin-1/2 nucleon have been investigated extensively for finding the origin of the nucleon spin in terms of quarks and gluons.
However, polarized structure functions of spin-1 hadrons have not been well studied experimentally at high energies.
In the spin-1 hadrons, there are structure functions which do not exist in the spin-1/2 nucleon~\cite{sk-2024-spin-1}.
For example, tensor-polarized structure functions and the gluon transversity exist as new observables for hadrons with spin$\,\ge 1$.

As a stable spin-1 nucleus, we may use the deuteron as a target in lepton deep inelastic scattering (DIS).
There was a HERMES measurement on the tensor-polarized structure function $b_1$ in 2005~\cite{hermes-2005}.
Since then, there were a proposal and a letter of intent at the Thomas Jefferson National Accelerator Facility (JLab) on $b_1$~\cite{Jlab-b1} and the gluon transversity~\cite{jlab-gluon-trans,gluon-trans-1}, respectively.
Actual experiments have not been done after 2005; however, the situation is rapidly changing.
Now, the $b_1$ proposal is fully approved to start the experiment.
Therefore, the field of spin-1 structure functions could become one of the exciting topics in hadron physics in the near future.
In addition, the spin-1 structure functions could be studied at the Fermi National Accelerator Laboratory (Fermilab)~\cite{ks-tensor-DY-2016,ks-trans-g-2020,Fermilab-spin,Keller-2022}, the Nuclotron-based Ion Collider fAcility (NICA)~\cite{NICA-2021}, the Large Hadron Collider (LHC)~\cite{LHC-spin}, and the Electron-Ion Colliders (EICs)~\cite{eic-2022,EicC-2021}.

For the spin-1 deuteron, there are four new polarized structure functions $b_{1\text{-}4}$~\cite{spin-1-deuteron-sfs}, which do not exist in the spin-1/2 nucleon, in the charged-lepton DIS.
For the twist-2 function $b_1$, there exists a sum rule~\cite{b1sum}, which was tested in the HERMES experiment~\cite{hermes-2005}. 
However, it seems that the $b_1$ values of the HERMES are very different from standard deuteron model calculations based on a convolution model~\cite{b1-convolution-2017}.
It suggests that a new hadronic mechanism may be needed for explaining the HERMES data~\cite{b1-miller-2014}.
Another interesting observable of a spin-1 hadron is the gluon transversity $\Delta_T g$, which does not exist in the nucleon, and it could also be measured by future experiments at lepton and hadron accelerator facilities
~\cite{jlab-gluon-trans,ks-trans-g-2020,Keller-2022,NICA-2021,eic-2022,EicC-2021}.

On the other hand, the twist-2 PDFs, transverse-momentum-dependent parton distributions (TMDs), and fragmentation functions (FFs) for spin-1 hadrons were investigated in 2000~\cite{bm-2000}, and twist-3 and 4 functions were clarified in 2021~\cite{ks-tmd-2021}.
Then, for these higher-twist distributions, useful relations were obtained in a similar way to the Wandzura-Wilczek (WW) relation and the Burkhardt-Cottingham (BC) sum rule for spin-1/2 hadrons~\cite{twist-2-relations-ks-2021} and also by using the equation of motion for a quark~\cite{eq-motion-ks-2021}.
In addition, generalized parton distributions (GPDs) for spin-1 hadrons were studied in Ref.~\cite{GPDs-spin-1}, and there were lightcone-model studies on the spin-1 $\rho$-meson GPDs~\cite{GPDs-rho}, and TMDs~\cite{TMDs-rho}.

The JLab experiments for the tensor-polarized deuteron will be done in the near future.
$Q^2 (=-q^2)$ values, where $q$ is the momentum transfer in the electron scattering, are not much larger than the typical hadronic energy scale of the order of 1~GeV$^2$ in the JLab measurements.
This fact suggests that higher-twist terms should be understood in extracting even the twist-2 tensor-polarized structure functions.
Now, the theoretical formalism is completed to calculate cross sections of inclusive and semi-inclusive DIS (SIDIS) processes for spin-1 hadrons~\cite{spin-1-SDIS}, and these processes could be used for measuring the tensor-polarized twist-3 distribution functions $f_{LT}(x)$.
In addition, Drell-Yan processes could be used for studying $f_{LT}(x)$ at the hadron accelerator facilities~\cite{Song-twist3-2025}.
Considering this situation, we show possible twist-3 distribution functions $f_{LT}(x)$ theoretically for the deuteron in this work.
Because the distributions $f_{LT}(x)$ are written in terms of the twist-2 distributions $f_{1LL}(x)$ by the WW-like relation and a parametrization exists for $f_{1LL}(x)$, actually tensor-polarized PDFs $\delta_T q (x)$~\cite{tensor-pdfs}, we can calculate $f_{LT}(x)$ by using this twist-2 relation.
The obtained $f_{LT}(x)$ should satisfy the BC-like sum rule.

This article consists of the following.
In Sec.~\ref{spin-1-sfs}, the tensor-polarized PDFs of spin-1 hadrons are introduced for explaining $f_{LT}(x)$ and $f_{1LL}(x)$.
Numerical results are shown in Sec.~\ref{results}, and they are summarized in Sec.~\ref{summary}.

\section{Twist-2 and 3 tensor-polarized parton distribution functions of the spin-1 deuteron}
\label{spin-1-sfs}

The PDFs are defined from correlation functions for the spin-1 deuteron.
The correlation function is given by the amplitude to extract a parton from the deuteron and then to insert it into the deuteron at a different spacetime point $\xi$, and it is expressed as
\begin{equation}
    \label{eqn:correlation-q}
    \Phi_{ij}^{[c]} (k, P, T)  
    = \! \! \int \! \frac{d^4 \xi}{(2\pi)^4} \, e^{ i k \cdot \xi} \langle  P, T \left | \, \bar\psi _j (0)  W^{[c]} (0, \xi) \psi _i (\xi)  \, \right | P, T  \rangle .
\end{equation} 
Here, $k$, $P$, and $T$ are the quark momentum, the deuteron momentum, and the tensor polarization of the deuteron, $\psi$ is the quark field, and $W^{[c]}$ is the gauge link with the integral path $c$ necessary for the color gauge invariance.
The tensor $T^{\mu\nu}$ is symmetric and traceless, and it satisfies $P_\mu T^{\mu\nu}=0$.
Then, integrating the function over the momenta $k^-$ and the transverse momentum $\vec k_T$ and fixing the $k^+$ component, we obtain the collinear correlation function
\begin{multline}
    \label{eqn:correlation-pdf}
    \Phi_{ij} (x_{D}, P, T ) 
    = \int d^2 k_T \, dk^+ dk^- \, \Phi^{[c]}_{ij} (k, P, T ) \, \delta (k^+  -x_{D} P^+) \\
    = \int \! \frac{d\xi^-}{2\pi} \, e^{i x_{D} P^+ \xi^-} \langle \, P , T \left | \, \bar\psi _j (0)  W^{[c]} (0, \xi) \psi _i (\xi)  \, \right |  P, \,  T \, \rangle _{\xi^+ =0, \, \vec\xi_T=0} ,
\end{multline}
where $k^\pm =(k^0 \pm k^3)/\sqrt{2}$ and the momentum fraction $x_{D}$ is defined by $k^+ = x_{D} P^+$.
The notation $D$ is used by considering the deuteron, but the expressions of this Sec.~\ref{spin-1-sfs} are valid for any spin-1 hadrons by denoting $x_{D}$ as $x_h$, where $h$ is a hadron, or simply as $x$.
However, the notation $x$ is used in Sec.~\ref{results} of this paper for a different variable, so that $x_{D}$ 
is used throughout this section.
This $x_{D}$ corresponds to $x$ in Refs.~\cite{ks-tmd-2021,twist-2-relations-ks-2021,eq-motion-ks-2021} which are written for any spin-1 hadrons.

For the spin-1 deuteron, the spin tensor $T^{\mu\nu}$ is parametrized
as~\cite{bm-2000,ks-tmd-2021}
\begin{align}
& \hspace{-0.20cm}
T^{\mu\nu} = \frac{1}{2} \left [ \frac{4}{3} S_{LL} \frac{(P\cdot n)^2}{M_D^2} 
               \bar n^\mu \bar n^\nu 
          - \frac{2}{3} S_{LL} ( \bar n^{\{ \mu} n^{\nu \}} -g_T^{\mu\nu} )
\right.
\nonumber \\
& \hspace{-0.35cm}
\left.
+ \frac{1}{3} S_{LL} \frac{M_D^2}{(P\cdot n)^2}n^\mu n^\nu
+ \frac{P \cdot n}{M_D} \bar n^{\{ \mu} S_{LT}^{\nu \}}
- \frac{M_D}{2 P \cdot n} n^{\{ \mu} S_{LT}^{\nu \}}
+ S_{TT}^{\mu\nu} \right ],
\label{eqn:spin-1-tensor-1}
\end{align}
where $M_D$ is the deuteron mass and
$a^{\{ \mu} b^{\nu \}}$ is defined by
$a^{\{ \mu} b^{\nu \}} = a^\mu b^\nu + a^\nu b^\mu$.
The lightcone vectors $n$ and $\bar n$ are defined by
\begin{equation}
    \label{eqn:lightcone-n-nbar}
    n^\mu =\frac{1}{\sqrt{2}} (\, 1,\, 0,\, 0,\,  -1 \, ), \ \ \
    \bar n^\mu =\frac{1}{\sqrt{2}} (\, 1,\, 0,\, 0,\,  1 \, ) .
\end{equation}
The $g_T^{\mu\nu}$ is the transverse projector,
$g_T^{\mu\nu} = g^{\mu\nu} - \bar n^{\{ \mu} n^{\nu\}}$.
The tensor polarization along the longitudinal axis is expressed by $S_{LL}$.
The parameters $S_{LT}^{\mu}$ and $S_{TT}^{\mu\nu}$ indicate 
polarization differences along the axes 
between the longitudinal and transverse directions 
and along the transverse axes, respectively~\cite{bm-2000,ks-tmd-2021}.
They satisfy the relations
$S_{LT} \cdot \bar{n} = S_{LT} \cdot n = 0$,
$S_{TT}^{\mu\nu} \bar{n}_\mu = S_{TT}^{\mu\nu} n_\mu = 0$ and
${g_T}_{\mu\nu} S_{TT}^{\mu\nu} = 0$.

The tensor-polarized part of the collinear correlation function is expressed by four PDFs $f_{1LL}$, $e_{LL}$, $f_{LT}$, and $ f_{3LL}$ up to twist 4 with the tensor-polarization parameters $S_{LL}$ and $S_{LT}$ as~\cite{twist-2-relations-ks-2021}
\begin{align}
    \label{eqn:collinear-correlation-pdfs}
    \Phi (x_{D}) 
    = & \frac{1}{2} \, \bigg[ \, S_{LL} \, \slashed{\bar n} \, f_{1LL} (x_{D}) + \frac{M_D}{P \cdot n} \, S_{LL} \, e_{LL} (x_{D}) \nonumber \\
    & + \frac{M_D}{P \cdot n} \, \slashed{S}_{LT} \, f_{LT} (x_{D}) + \frac{M_D^2}{(P \cdot n)^2} \, S_{LL} \, \slashed{n} \, f_{3LL} (x_{D}) \, \bigg] ,
\end{align}
where $\slashed{a} = \gamma^{\,\mu} a_\mu$.
The quark distribution function $f_{1LL} (x_{D})$ is a twist-2 function,
the functions $e_{LL} (x_{D})$ and $f_{LT} (x_{D})$ are twist-3 ones,
and the function $f_{3LL} (x_{D})$ is a twist-4 one.
In this work, we try to obtain a possible distribution $f_{LT} (x_{D})$
from $f_{1LL} (x_{D})$.

In the polarized structure functions of the nucleon, the WW relation 
has been often used for estimating the twist-3 structure function $g_2$ 
in terms of the twist-2 one $g_1$.
In addition, the BC sum rule is also useful for the estimate of $g_2$.
For a review of the original WW relation and the BC sum rule in the spin-1/2 case, 
one may look at Ref.~\cite{kt-1999}.
In a similar way to the derivations of these relations, 
the WW-like relation was obtained for the distribution functions $f_{LT} (x_{D})$ 
and $f_{1LL} (x_{D})$ by neglecting dynamical twist-3 terms as
~\cite{twist-2-relations-ks-2021}
\begin{equation}
    \label{eqn:flt-twist2}
    f_{LT}^{\, q+}(x_D)= \frac{3}{2} \int^1_{x_D} \frac{dy_D}{y_D} \, f_{1LL}^{\, q+}(y_D) ,
\end{equation}
and the BC-like sum rule was derived
\begin{equation}
    \label{eqn:f_lt-sum}
    \int_0^1 dx_D \, f_{2LT}^{\, q+} (x_D) =0 , \ \ \ 
    f_{2LT}^{\, q+} (x) \equiv \frac{2}{3} f_{LT}^{\, q+} (x_D) - f_{1LL}^{\, q+} (x_D) .
\end{equation}
Here, the quark flavor type is shown by $q$ in the superscript as $f^{\, q+}$, which is defined by
\begin{equation}
    \label{eqn:quark-antiquark}
    f^{\, q+} (x_D) \equiv f^{\, q} (x_D) + f^{\, \bar q} (x_D), \ \ 
    f=f_{1LL},\ f_{LT},\ f_{2LT} .
\end{equation}
In this way, if we have the knowledge of the twist-2 distributions $f_{1LL}^{\, q+} (x_D)$, the twist-3 distributions $f_{LT}^{\, q+} (x_D)$ can be calculated numerically.

The  original WW relation was derived in terms of 
the operator product expansion (OPE) with local operators~\cite{kt-1999,ope-local}. 
In our case, the WW-like relation was obtained 
by nonlocal operators with the tensor polarization 
parameters~\cite{twist-2-relations-ks-2021}.
It should be noted that the same WW-like relation is obtained also
by the local OPE with the mixed-symmetry operator 
$\bar \psi (0) \gamma^{\, [ \sigma} D^{\, \{ \mu_1 ]} \cdots D^{\, \mu_{n-1} \}} \psi (0)$
and the symmetry one 
$\bar \psi (0) \gamma^{\, \{ \sigma} D^{\, \mu_1} \cdots D^{\, \mu_{n-1} \}} \psi (0)$,
where $[\cdots]$ indicates the antisymmetric combination.
Their matrix elements are decomposed into the mixed-symmetric and symmetric terms
in terms of the momentum $P^{\, \mu}$ and the tensor $T^{\mu\nu}$. 
Then, the WW-like relation is obtained in a similar way 
as explained in Sec.~2 of Ref.~\cite{twist-2-relations-ks-2021}.
Therefore, the current WW-like relation is consistent 
with the local OPE derivation.

\section{Results on $f_{LT}$ for the deuteron}
\label{results}

In order to calculate Eq.~\eqref{eqn:flt-twist2}, we need an appropriate functions $f_{1LL}^{\, q+}$ on the right-hand side.
One may be careful in using the relations in Sec.~\ref{spin-1-sfs}, because a different variable $x$ instead of $x_D$ is usually used for expressing the nuclear structure functions and PDFs. 
For example, the Bjorken scaling variables could be defined as
\begin{equation}
    \label{eqn:x-Bjorken}
    x = \frac{Q^2}{2 M_N \nu}, \ \ \ 
    x_{D} = \frac{Q^2}{2 P \cdot q} = \frac{Q^2}{2 M_D \nu} ,
\end{equation}
where $M_N$ is the nucleon mass and $\nu \, (=q^0$) is the energy transfer in the lepton DIS process with the fixed deuteron target.
They are defined in the ranges
\begin{equation}
    \label{eqn:x-ranges}
    0 \le x \le 2, \ \ \ 
    0 \le x_{D} \le 1 .
\end{equation}
In expressing nuclear structure functions and PDFs, this scaling variable $x$ is usually used, which is sometimes confusing.
In this section, the variable $x$ of Eq.~\eqref{eqn:x-Bjorken} is used, so that its range is $0 \le x \le 2$.

In the following, the used distributions $f_{1LL}^{\, q+} (x)$ are explained first in Sec.~\ref{f1ll}, and the distributions $f_{LT}^{\, q+} (x)$ are shown numerically in Sec.~\ref{flt} by calculating the WW-like relation.
Then, we confirm that the obtained distributions satisfy the BC-like sum rule.

\subsection{Twist-2 distribution functions $f_{1LL}(x)$}
\label{f1ll}

So far, we have been using the Trento-type convention for the tensor-polarized PDFs $f_{1LL}$, $f_{LT}$, and $f_{3LL}$.
The quark distribution $f_{1LL}^{\, q}$, where the quark flavor $q$ is explicitly written, is also expressed by the notation $\delta_T q$ ($=b_1^q$). 
In Ref.~\cite{tensor-pdfs}, the tensor-polarized PDFs were determined to explain the HERMES data on $b_1$ for the deuteron at $Q^2=2.5$~GeV$^2$, which is the average $Q^2$ of the HERMES data.
The tensor-polarized PDFs of the deuteron ($D$) are expressed for the valence quarks and antiquarks as
\begin{equation}
    \label{eqn:delta-q-qbar-1}
    \delta_{_T} q_{v}^{\text{\tiny $D$}} (x) = \delta_{_T} w(x) \, q_{v}^{\text{\tiny $D$}} (x), \ \ 
    \delta_{_T} \bar q^{\text{\tiny $D$}} (x) = \alpha_{\bar q} \, \delta_{_T} w(x) \, \bar q^{\text{\tiny $D$}} (x) ,
\end{equation}
where $q_{iv}^{\text{\tiny $D$}} (x)$ and $\bar q_i^{\text{\tiny $D$}} (x)$ are the unpolarized PDFs of the deuteron per the nucleon, namely the deuteron PDFs divided by 2, the function $\delta_{_T} w(x)$ indicates the fraction for the tensor polarization, and $\alpha_{\bar q}$ is the extra parameter for the antiquark distribution.
The tensor-polarized PDF $\delta_{_T} q^D$ is defined as
\begin{equation}
    \label{eqn:delta_Tq}
    \delta_{_T} q^D \equiv q^{D \, 0} - \frac{q^{D \, +1}+q^{D \, -1}}{2} ,
\end{equation}
where $q^{D \, \lambda}$ indicates a quark distribution with the flavor $q$ in the deuteron spin state $\lambda$.
Sometimes, the notation $b_1^{q/D}$ is used instead of $\delta_{_T} q^D$ ($b_1^{q/D}=\delta_{_T} q^D$) for the tensor-polarized PDFs.

These PDFs are given at $Q^2=2.5$~GeV$^2$, where the charm- and bottom-quark distributions are neglected, and the gluon distribution $\delta_T g^D$ is assumed to be zero at this $Q^2$ scale. 
Furthermore, the flavor dependence of the tensor-polarized PDFs is not considered due to a lack of data at this stage:
\begin{align}
    \label{eqn:dw(x)}
    \delta_{_T} q_v^{\text{\tiny $D$}} (x)
    & \equiv \delta_{_T} u_v^{\text{\tiny $D$}} (x) 
    = \delta_{_T} d_v^{\text{\tiny $D$}} (x) 
    = \delta_{_T} w(x) \, \frac{u_v (x) +d_v (x)}{2} , 
    \nonumber \\
    \delta_{_T} \bar q^{\text{\tiny $D$}} (x)
    & \equiv \delta_{_T} \bar u^{\text{\tiny $D$}} (x)
    = \delta_{_T} \bar d^{\text{\tiny $D$}} (x)
    = \delta_{_T}      s^{\text{\tiny $D$}} (x)
    = \delta_{_T} \bar s^{\text{\tiny $D$}} (x)                    
    \nonumber \\
    & = \alpha_{\bar q} \, \delta_{_T} w(x) \, \frac{2 \bar u(x) +2 \bar d(x) +s(x) + \bar s(x)}{6} .
\end{align}
Here, the PDFs on the right-hand sides without the superscript $D$, namely $u_v$, $d_v$, $\bar u$, $\bar d$, $s$, and $\bar s$, are the functions of the nucleon.
This assumption leads to the restriction that the tensor-polarized PDFs of the deuteron are zero in the range $1 \le x \le 2$, whereas finite distributions could exist in this large-$x$ region as given, for example, in Ref.~\cite{b1-convolution-2017}, although the distributions are small at $x>1$.

\begin{figure}[b]
    \centering
    \includegraphics[width=0.85\linewidth]{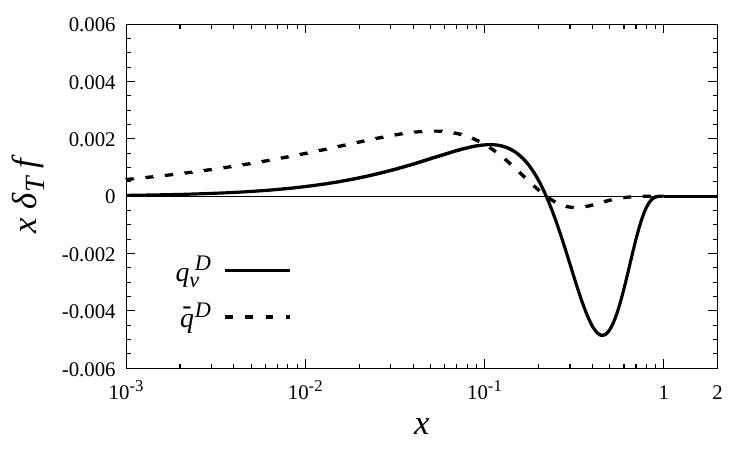}
    \vspace{-0.4cm}
    \caption{Tensor polarized PDFs $\delta_T f$ for the deuteron at $Q^2=2.5$~GeV$^2$ for explaining the HERMES data. The PDFs are shown by the ones per nucleon.}
    \label{fig:delta_Tf-x-2025}
\end{figure}

From these PDFs, the structure function $b_1^D(x)$ is given as $b_1^D (x)  = \frac{1}{2} \sum_q e_q^2 \, \left [ \delta_{_T} q^D (x) + \delta_{_T} \bar q^D (x)   \right ] $ with the quark charge $e_q$,
and the functional form of $\delta_{_T} w(x) $ is chosen so that the sum $\int_0^2 dx \, b_1^D (x) = 0$~\cite{b1sum} is satisfied:
\begin{equation}
    \label{eqn:dw(x)-abc}
    \delta_{_T} w(x) = a x^b (1-x)^c (x_0-x) .
\end{equation}
The factor $1/2$ in $b_1^D (x)$ is because it is defined by the one per nucleon.
Here, $a$, $b$, $c$, and $x_0$ are the parameters.
The set-2 analysis parameters~\cite{tensor-pdfs} are used in this work:
\begin{equation}
    \label{eqn:parameter-values}
    a=0.221,\ \ \alpha_{\bar q}=3.20,\ \ b=0.648,\ \ c=1,\ \ x_0=0.221 .
\end{equation}
The tensor-polarized PDFs obtained by these parameters are shown in Fig.~\ref{fig:delta_Tf-x-2025}.
Because of the $x$-dependent functional form defined in Eq.~\eqref{eqn:dw(x)} with Eq.~\eqref{eqn:dw(x)-abc} to satisfy the integral sum, the distributions have oscillatory behavior.

\subsection{Twist-3 distribution functions \texorpdfstring{$f_{LT} (x)$}{flt(x)} by using the twist-2 relations}
\label{flt}

\begin{figure}[b]
    \centering
    \includegraphics[width=0.85\linewidth]{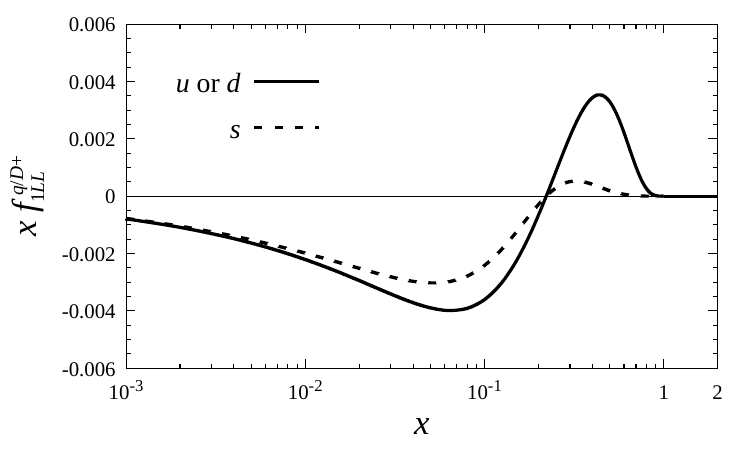}
    \vspace{-0.4cm}
    \caption{Tensor-polarized twist-2 distribution functions $xf_{1LL}^{\, q/D +} = x f_{1LL}^{\, q/D} + x f_{1LL}^{\, \bar q/D}$ are shown for the deuteron at $Q^2=2.5$~GeV$^2$. The solid curve indicates the distribution for $q=u$ or $d$, and the dashed curve is the distribution for $q=s$. The PDFs are shown by the ones per nucleon.}
    \label{fig:f_1LL-x-2025}
\end{figure}
\begin{figure}[b!]
    \centering
    \includegraphics[width=0.85\linewidth]{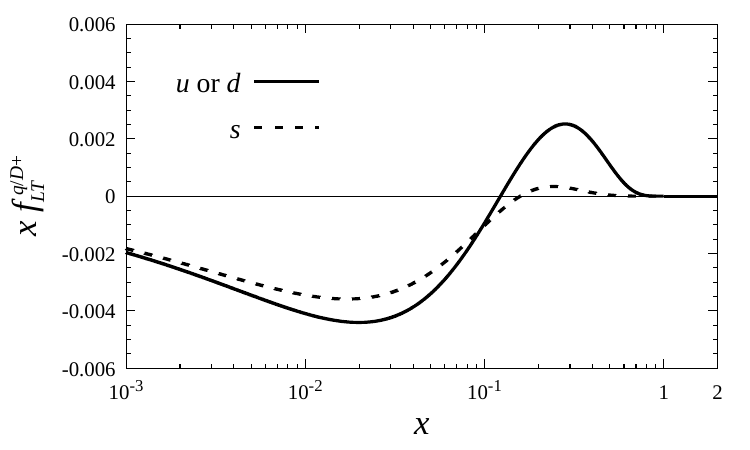}
    \vspace{-0.4cm}
    \caption{Tensor-polarized twist-3 distribution functions $xf_{LT}^{\, q/D +} = x f_{LT}^{\, q/D} + x f_{LT}^{\, \bar q/D}$ are shown for the deuteron at $Q^2=2.5$~GeV$^2$. The solid curve indicates the distribution for $q=u$ or $d$, and the dashed curve is the distribution for $q=s$. The PDFs are shown by the ones per nucleon.}
    \label{fig:f_LT-x-2025}
\end{figure}

In Sec.~\ref{spin-1-sfs}, the twist-2 relations are expressed by the variable $x_D$ in the range up to $1$.
However, the tensor-polarized PDFs of Sec.~\ref{f1ll} are expressed by the variable $x$ up to $2$, so that
the twist-2 relations of Eqs.~\eqref{eqn:flt-twist2} and \eqref{eqn:f_lt-sum} are re-written as
\begin{equation}
    \label{eqn:flt-twist2-mod}
    f_{LT}^{\, q/D+}(x)= \frac{3}{2} \int^2_x \frac{dy}{y} \, f_{1LL}^{\, q/D+}(y) ,
\end{equation}
for the WW-like relation and
\begin{equation}
    \label{eqn:f_lt-sum-mod}
    \int_0^2 dx \, f_{2LT}^{\, q/D+} (x) =0 , \ \ 
    f_{2LT}^{q/D+} (x) \equiv \frac{2}{3} f_{LT}^{\, q/D+} (x) - f_{1LL}^{\, q/D+} (x) ,
\end{equation}
for the BC-like sum rule by using the variable $0 \le x \le 2$.
Here, the distribution is redefined as $f_{\text{Sec.2}}(x_D) = f_{\text{Sec.3}}(x/2) \equiv f(x)$.

Using the PDFs in Fig.~\ref{fig:delta_Tf-x-2025}, we calculated the tensor-polarized twist-2 distribution 
functions $f_{1LL}^{\, q/D+}$, which are expressed by the PDFs $\delta_T q^D$ and $\delta_T \bar q^D$ as~\cite{sk-2024-spin-1,twist-2-relations-ks-2021}
\begin{align}
    \label{eqn:f_1LL-delta_Tq}
    f_{1LL}^{\, q/D+} (x) & = f_{1LL}^{\, q/D} (x) + f_{1LL}^{\, \bar q/D} (x)
    = -\frac{2}{3} \left[ \delta_T q^D (x) + \delta_T \bar q^D (x) \right ] \nonumber \\
    & = -\frac{2}{3} \left[ \delta_T q_v^D (x) + 2 \delta_T \bar q^D (x) \right ],      
\end{align}
where the distributions are separately written by the valence-quark distributions $\delta_T q_v^D (x)$ and the antiquark ones $\delta_T \bar q^D (x)$.
We show the distributions $f_{1LL}^{\, q/D+} (x)$ at $Q^2=2.5$~GeV$^2$.
Because of the relations in Eq.~\eqref{eqn:dw(x)}, the $u$ and $d$ quark distributions are the same: $f_{1LL}^{\, u/D+} (x) = f_{1LL}^{\, d/D+} (x)$, and the strange-quark distribution $f_{1LL}^{\, s/D+} (x)$ is given by Eq.~\eqref{eqn:f_1LL-delta_Tq} without the valence-quark term  $\delta_T s_v^D (x)=0$.

Obtained distributions $x f_{1LL}^{\, u/D+} (x)=x f_{1LL}^{\, d/D+} (x)$ and $x f_{1LL}^{\, s/D+} (x)$ are shown at $Q^2 =2.5$~GeV$^2$ in Fig.~\ref{fig:f_1LL-x-2025} as the function of $x$.
Because of the overall sign change in Eq.~\eqref{eqn:f_1LL-delta_Tq}, the distributions are positive at $x>0.2$ and they are negative at small $x \, (<0.2)$. 
Integrating these distributions by Eq.~\eqref{eqn:flt-twist2-mod}, we obtain the twist-3 distributions 
$x f_{LT}^{\, u/D+} (x)=x f_{LT}^{\, d/D+} (x)$ and $x f_{LT}^{\, s/D+} (x)$ at $Q^2=2.5$~GeV$^2$ in Fig.~\ref{fig:f_LT-x-2025}.
The function forms and magnitudes of these distributions are roughly similar to those of $x f_{1LL}^{\, u/D+} (x)$ and $x f_{1LL}^{\, s/D+} (x)$.

\begin{figure}[t]
    \centering
    \includegraphics[width=0.85\linewidth]{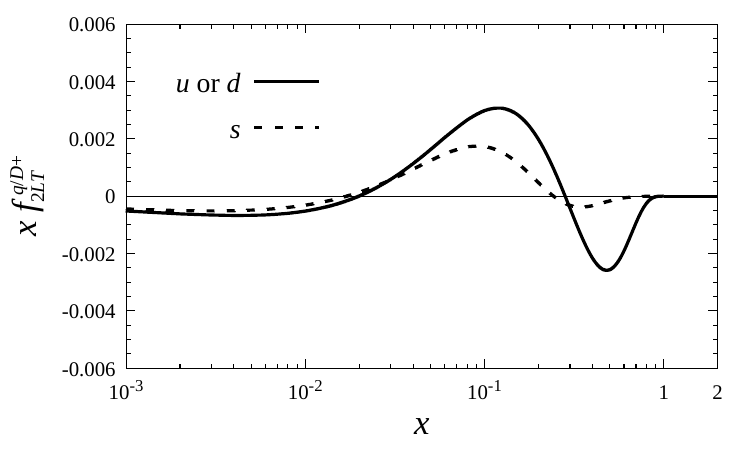}
    \vspace{-0.4cm}
    \caption{Tensor-polarized twist-3 distribution functions $xf_{2LT}^{\, q/D +} = x f_{2LT}^{\, q/D} + x f_{2LT}^{\, \bar q/D}$ are shown for the deuteron at $Q^2=2.5$~GeV$^2$. The solid curve indicates the distribution for $q=u$ or $d$, and the dashed curve is the distribution for $q=s$. The PDFs are shown by the ones per nucleon.}
    \label{fig:f_2LT-x-2025}
\end{figure}

In order to confirm the BC-like sum rule in Eq.~\eqref{eqn:f_lt-sum-mod}, the distributions $f_{1LL}^{\, q/D+} (x)$ and $f_{LT}^{\, q/D+} (x)$ are combined to obtain to $f_{2LT}^{\, q/D+} (x)$ at $Q^2=2.5$~GeV$^2$ in Fig.~\ref{fig:f_2LT-x-2025}. According to the sum rule, the integrals of these distributions should vanish.
In fact, the positive and negative areas in this figure cancel with each other.
However, finite distributions still exist at $x \le 10^{-3}$ in Fig.~\ref{fig:f_2LT-x-2025} and the integrals become much smaller, of the order of $10^{-5}$, if the integrals are calculated from $x_{\min}=10^{-9}$.

The higher-twist distributions are generally suppressed by $1/Q$ in cross sections in comparison with 
the leading-twist ones.
However, the $Q^2$ values of the JLab experiment are not very large, so that there is a possibility
that the higher-twist distributions $f_{LT}^{\, q/D+} (x)$ could be measured experimentally by the 
inclusive and semi-inclusive DIS (SIDIS)~\cite{spin-1-SDIS} 
in addition to the leading-twist ones $f_{1LL}^{\, q/D+} (x)$~\cite{Jlab-b1}.
The SIDIS cross section is expressed by various functions denoted as~\cite{spin-1-SDIS}
\vspace{-0.15cm}
\begin{equation}
F_\text{incoming lepton spin (target spin), photon polarization}^\text{\ azimuthal angle} .
\nonumber
\vspace{-0.15cm}
\end{equation}
The twist-3 distribution function $f_{LT}^{\, q/D+} (x)$ is contained in the function
$F_{U(LT)}^{\, \cos \phi_{LT}}$. 
Namely, using the unpolarized charged-lepton beam and the target with the tensor polarization (LT)
and looking at the dependence of $\cos \phi_{LT}$ in the cross section, 
we can measure the function $f_{LT}^{\, q/D+} (x)$.
Here, $\phi_{LT}$ is the azimuthal angle of tensor polarization $S_{LT}^\mu$.
According to Eqs.~(4.111) and (4.113)-(4.116) in Ref.~\cite{spin-1-SDIS}, 
$f_{LT}^{\, q/D+} (x)$ is contained in the structure functions $b_{2\text{--}4}$ in inclusive DIS.
Thus, the distribution $f_{LT}^{\, q/D+} (x)$ could be measured at JLab \cite{Jlab-b1,jlab-gluon-trans} 
and future EICs \cite{eic-2022,EicC-2021}. 

Furthermore, Drell-Yan processes could also be used for finding the twist-3 distributions $f_{LT}^{\, q/D+} (x)$~\cite{Song-twist3-2025}.
In the proton-deuteron Drell-Yan process with the tensor-polarized deuteron,
the function $f_{LT}^{\,q/D+} (x)$ appears in the cross section with
the angular dependence of $\cos \hat\phi$, where $\hat\phi$ is 
$\hat\phi = \phi-\phi_s$ with the azimuthal angle $\phi$ of $\mu^-$ 
and the transverse-vector angle $\phi_s$
of $S_{LT}^\mu$.
Therefore, there is a possibility that $f_{LT}^{\,q/D+} (x)$ could be measured
in the Fermilab experiment \cite{Fermilab-spin,Keller-2022} and 
in other experiments at the hadron accelerator facilities \cite{NICA-2021,LHC-spin}.   

\section{Summary}
\label{summary}

Using the parametrization for the tensor-polarized PDFs $\delta_T q^D (x)$ and $\delta_T \bar q^D (x)$, we calculated the tensor-polarized twist-3 PDFs $f_{LT}^{\, u/D +} (x)=f_{LT}^{\, d/D +} (x)$ and $f_{LT}^{\, s/D +} (x)$ for the deuteron at $Q^2=2.5$~GeV$^2$ by using the twist-2 Wandzura-Wilczek-like relation.
This is the first numerical estimate of the tensor-polarized twist-3 distributions.
Their functional forms and magnitudes are roughly similar to the twist-2 distributions $f_{1LL}^{\, u/D +} (x)=f_{1LL}^{\, d/D +} (x)$ and $f_{1LL}^{\, s/D +} (x)$.
We also confirmed that the Burkhardt-Cottingham-like sum rule is satisfied numerically.
The twist-3 distributions $f_{LT}^{\, q/D +} (x)$ should be investigated for the deuteron by the future inclusive DIS, semi-inclusive DIS, and Drell-Yan measurements.

\section*{Acknowledgments}

The authors were partially supported by a research program 
of the Chinese Academy of Sciences.
KK was also supported by the Gansu-province postdoctoral foundation.
They thank Qin-Tao Song for suggestions.


\end{document}